# Requisitos para um rádio digital interativo no Brasil e América Latina

Rafael Diniz
Departamento de Informática
PUC-Rio
Rua Marquês de São Vicente, 225
Rio de Janeiro/RJ – Brasil
+55 21 35271500
rdiniz@inf.puc-rio.br

## RESUMO

Este artigo discute a viabilidade e os requisitos para a utilização do Ginga como o middleware de um Sistema de Rádio Digital. O Ginga foi adotado pelo Brasil e vários outros países da América Latina como a norma para a interatividade na TV Digital. Neste artigo os dois padrões de rádio digital sendo considerados para adoção pelo Brasil serão brevemente apresentados, uma discussão sobre os requisitos necessários para a interatividade no rádio é feita e finalmente um protótipo de um aplicativo Ginga para rádio digital que foi transmitido e recebido via rádio é detalhado.

## INTRODUÇÃO

Iniciando com um breve histórico sobre o rádio digital no Brasil, em 2010 o Sistema Brasileiro de Rádio Digital (SBRD) foi instituído pela Portaria nº 290/2010[1] do Ministério das Comunicações, assinada pelo então ministro Hélio Costa. Essa portaria não definiu uma norma técnica para o sistema, dando somente diretrizes de alto nível com características que SBRD deverá ter[2].

Em 2011 o Ministério das Comunicações publicou o chamamento público número 1/2011[3] que convocou sistemas de rádio digital a se apresentarem para testes e avaliações por parte do governo brasileiro, com o propósito de definir um o padrão técnico para o SBRD. Dois sistemas de rádio digital demonstraram interesse em participar do processo: o HD Radio e o Digital Radio Mondiale. Na Seção 1 deste artigo ambos padrões são apresentados.

Em 2012 o Ministério das Comunicações, através da Portaria nº 365/12[4] criou o Conselho Consultivo do Rádio Digital com o objetivo de assessorar o Ministro de Estado das Comunicações na implantação do Rádio Digital no Brasil. O Conselho conta com a presença de representantes do setor industrial, da radiodifusão e do governo. A academia brasileira não foi convidada a participar.

Entre 2010 a 2012 foram realizadas 11 baterias de testes em 7 emissoras com os dois padrões de rádio digital.

O Ginga, padrão de interatividade do Sistema Brasileiro de TV Digital, parece ser uma opção natural ao se pensar em interatividade no rádio digital, principalmente devido ao fato do Ginga já estar presente em receptores fixos e móveis de TV, como smartphones, tablets e GPS que possuem receptor de TV Digital One-Seg embutido, portanto sua adoção poderá ser quase tão simples quanto utilizar o mesmo middleware de interatividade tanto pelo receptor de TV quanto pelo receptor de rádio digitais embutidos em determinado dispositivo, consideradas as diferenças intrínsecas dos sistemas. Uma descrição do Ginga é feita na Seção 2.

De forma a testar a viabilidade e os requisitos técnicos para o uso do Ginga no rádio digital, uma discussão na Seção 3 do artigo é feita sobre esses requisitos necessários para a transmissão, recepção e execução de um aplicativo Ginga sobre um sistema de radiodifusão digital.

Na Seção 4 é detalhado um teste de um aplicativo Ginga que foi transmitido e recebido via um padrão de rádio digital e também executado com a implementação de referência do middleware Ginga.

Na última Seção é feita uma análise dos resultados obtidos e uma perspectiva para o futuro do rádio digital interativo no Brasil e quem sabe América do Sul e outros países.

## 1. OS PADRÕES EM CONSIDERAÇÃO PELO BRASIL

Dois padrões estão sendo considerados para adoção pelo Brasil de acordo com o Ministério das Comunicações[4]: O Digital Radio Mondiale (DRM) e o HD Radio. De forma simplificada, segue uma apresentação de ambos os sistemas.

O HD Radio é um padrão de rádio digital em uso principalmente nos Estados Unidos e México, e possui as seguintes características:

| Modulação | Orthogonal Frequency-Division Multiplexing |
|---|---|
| Codificador de áudio | HDC (proprietário / segredo industrial) |
| Modo de operação | Tido como um padrão hibrido, as portadoras digitais são posicionadas nos canais adjacentes superior e inferior ao sinal analógico AM ou FM. A distância em frequência do sinal digital para o analógico é fixa, sendo que a potência do sinal digital nunca pode ultrapassar um limiar que fica em torno de 10% da potência do sinal analógico, devido a interferência mútua dos sinais. O receptor chaveia entre o sinal digital e analógico caso o sinal digital se torne fraco. Ocupa 400kHz em VHF, já combinado com o sinal FM e 30kHz em OM, já combinado com o sinal AM. |
| Bandas de operação | Ondas Médias e VHF Banda II |
| Royalties / licença | Exige licença de uso da tecnologia para o rádio difusor, exige licença para integrar a tecnologia em transmissores e receptores[5], possui royalties mas o valores não são públicos. |
| Norma | "NRSC-5-B: In-band/on-channel Digital Radio Broadcasting Standard". Possui lacunas de informação como no caso do codificador de áudio que recebe somente uma menção na norma como "NOT SPECIFIED BY NRSC-5". Não existe documentação pública sobre o codificador / decodificador de áudio do HD Radio. |
| Responsável pelo sistema | O sistema é de propriedade de uma empresa de nome iBiquity Digital Corporation. No Brasil a empresa TellHD possui acordo com a Ibiquity e é a única empresa licenciadora da tecnologia do HD Radio no Brasil. |

O DRM é um padrão de rádio digital que foi desenvolvido por um consórcio de empresas e centros de pesquisa ligados ao rádio digital e está em uso principalmente na Índia e alguns países da Europa. Possui as seguintes características:

| Modulação | Orthogonal Frequency-Division Multiplexing |
|---|---|
| Codificador de áudio | HE AAC v2 (ISO/IEC 14496-3) |
| Modos de operação | É um sistema de rádio 100% digital, no qual o sinal digital é independente e não é relacionado à frequência de um sinal analógico, no entanto um link lógico pode ser feito entre as transmissões digitais e analógicas de forma a permitir a um receptor chavear de uma transmissão para outra, em caso de sinal fraco (utilizando a sinalização "Alternate Frequency Switching"). Pode funcionar em um canal adjacente (superior ou inferior) ao analógico, como em qualquer posição do espectro com relação ao sinal analógico e também no modo digital somente. Ocupa 100kHz em VHF, e pode ocupar 5kHz, 10kHz ou 20kHz nas frequências abaixo de 30MHz (OM, OT e OC). |
| Bandas de operação | OM, OT, OC e VHF bandas I, II e III |
| Royalties / licença | Não exige licença de uso da tecnologia para o rádio difusor, não exige licença para integrar a tecnologia em transmissores e receptores, possui royalties e os valores são públicos e cobrados por uma empresa especializada (Via Licensing) que de forma isonômica faz a arrecadação das taxas, cujos valores estão disponíveis publicamente[6]. |
| Norma | "ETSI ES 201 980: Digital Radio Mondiale (DRM); System Specification". Implementações da norma podem ser encontradas em software livre. |
| Responsável pelo sistema | O sistema é gerido pelo Consorcio DRM, que conta com importantes atores da radiodifusao e audio digital como Fraunhofer Institute, BBC, Deutsche Welle, Sony, Bosch, JVC Kenwood, All India Radio dentre outros. |

## 2. O MIDDLEWARE GINGA

O middleware Ginga é o sistema de interatividade que foi adotado pelo Sistema Brasileiro de TV Digital. Dois perfis do Ginga foram definidos para a TV Digital, um perfil para receptores móveis, e outro perfil para receptores fixos. Neste artigo iremos utilizar o perfil do Ginga para receptores móveis[7], e não será considerado para uso no rádio digital a extensão do Ginga presente somente em receptores de TV fixos, o Ginga-J, devido a importância da interoperabilidade entre todo tipo de receptor de TV e rádio digital. Todos os recursos presentes no perfil do Ginga para receptores móveis também estão presentes no perfil para receptores fixos.

Dentre outras características o Ginga é composto de uma linguagem declarativa XML que elenca os objetos de mídia do conteúdo interativo, seus relacionamentos e sincronismo, chamada de NCL (Nested Context Language). Além dessa linguagem declarativa é definido o objeto NCLua, que consiste de um objeto imperativo escrito na linguagem de programação LUA que pode interagir no contexto e nas diferentes propriedades dos objetos de mídia referenciados via NCL.

Vários tipos de mídia podem ser incluídos no aplicativo Ginga, como mídias de texto, imagem, áudio e vídeo.

A linguagem NCL foi concebida para ser utilizada por desenvolvedores de conteúdo multimídia e por isso tem uma sintaxe e semântica simples de ser compreendida por este perfil de desenvolvedor[8].

A adoção do Ginga no rádio digital em países que já utilizam o Ginga na TV Digital vai possibilitar a interoperabilidade dos dois meios. Um aplicativo Ginga proveniente de uma emissora de rádio, por exemplo, poderá referenciar um objeto de mídia transmitido por uma emissora de TV.

Além dessa vantagem o fato dos dispositivos receptores poderem compartilhar o mesmo middleware básico para ambos rádio e TV digitais irá favorecer a implementação de receptores que combinam rádio e TV digital interativos (como aparelhos celular). Uma outra característica do Ginga é o fato de sua implementação não implicar o pagamento de royalties.

## 3. REQUISITOS TÉCNICOS PARA O FUNCIONAMENTO DO GINGA NO RADIO DIGITAL

Num sistema de radiodifusão digital, pelo fato da informação transmitida serem bits, a priori, é possível mesclar na transmissão um aplicativo Ginga, que é composto de arquivos individuais com o código da aplicação e mídias associadas.

No entanto para esse tipo de transmissão conjunta ser plenamente possível, alguns requisitos devem estar disponíveis pelo sistema de rádio digital utilizado (ou deve-se desenvolver o suporte necessário), sendo os mais importantes:

- Um protocolo para a multiplexação da aplicação no fluxo de dados transmitido, sendo que o conteúdo seja transmitido em carrossel, de forma cíclica, e seja garantida a integridade dos dados enviados;
- Uma forma de sinalizar ao receptor o tipo de conteúdo que está sendo transmitido (no caso, um aplicativo Ginga), o ponto de entrada do aplicativo, se ele deve ser iniciado imediatamente ou não e um identificador único para o aplicativo;
- Pelo fato da estreita largura de banda de um canal de rádio, a taxa de bits útil que pode ser alocado para a transmissão de uma aplicação é pequeno, portanto a compactação dos dados da aplicação é muito relevante.

Considerados esses pré-requisitos como essenciais, faz-se necessário verificar a presença desses recursos em ambos os sistemas em análise pelo Brasil para o rádio digital.

No caso do DRM, é possível a utilização do protocolo MOT (Multimedia Object Transfer), que é definido no padrão do DRM e foi testado com sucesso, como será descrito na próxima seção. Uma apresentação durante a quarta reunião do Conselho Consultivo do Rádio Digital (CCRD) a respeito da viabilidade do Ginga no DRM foi feita[9].

No caso do HD Radio, foi entregue ao governo brasileiro na quinta reunião do CCRD um documento[10] redigido pela Ibiquity que propõe uma forma de integrar o Ginga ao HD Radio.

## 4. IMPLEMENTAÇÃO PROTÓTIPO DO GINGA SOBRE DRM

Para demonstrar a viabilidade da implementação do Ginga para o rádio digital foi desenvolvido um ambiente no qual foi realizada a transmissão e a recepção de um conteúdo Ginga utilizando-se o padrão DRM. A opção pelo DRM foi feita devido a disponibilidade do software de transmissão de forma gratuita e do software de recepção ser software livre, o que nos permitiu adaptá-lo de maneira que o mesmo chamasse o exibidor Ginga assim que a aplicação chegasse.

De forma a considerar o pior caso para a transmissão de um aplicativo interativo, no qual a taxa de bits disponível é muito baixa, escolhemos o modo de transmissão do DRM específico para frequências abaixo de 30MHz (DRM30). A frequência utilizada foi 12MHz, mas poderia ter sido qualquer frequência na faixa de OM, OT ou OC.

Do lado do transmissor foi utilizado o software Spark[11] de Michael Feilen (University of Applied Sciences in Kaiserslautern), o qual foi configurado para transmitir na faixa de Ondas Curtas, na frequência de 12MHz, tendo sido alocado 5kbps para envio do aplicativo Ginga e 16kbps para o áudio. O equipamento utilizado para transmitir o sinal foi a USRP[12]. O protocolo utilizado para transmitir o aplicativo Ginga e suas mídias foi o MOT[13], protocolo padronizado para envio de arquivos em carrossel para o DRM.

Do lado do receptor foi utilizado o receptor em software Dream[14], que suporta, dentre outros recursos, a decodificação do carrossel MOT. Para a aquisição do sinal foi utilizado o FUNcube Dongle Pro+[15]. Após o Dream retirar do carrossel os arquivos que compõem o aplicativo Ginga, o mesmo foi executado pelo exibidor Ginga[16] desenvolvido pela PUC-Rio.

Na Figura 1 está uma foto tirada no laboratório com do equipamento utilizado para a transmissão do sinal DRM, chamado USRP.

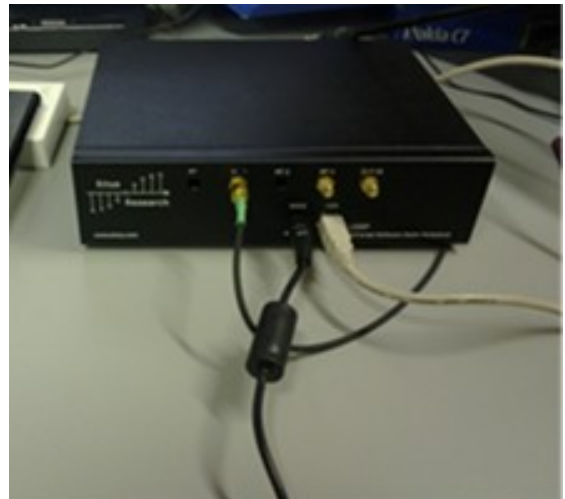

Fig. 1: Foto da USRP

O computador rodando o software Spark estava conectado à USRP e o computador rodando o Dream e o exibidor Ginga estava conectado ao FUN cube Dongle Pro+. O sinal foi transmitido a alguns centímetros de distância entre uma pequena antena conectada a USRP e um pequeno fio fazendo o papel de antena conectado ao FUN cube Dongle Pro+, que aparece na Figura 2.

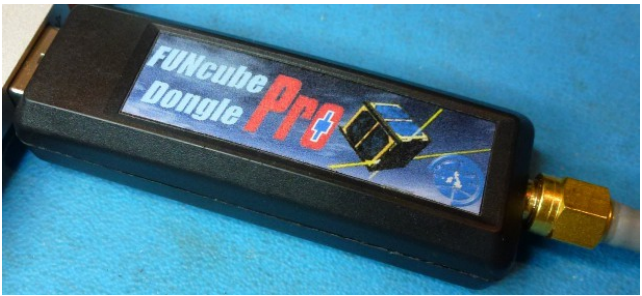

Fig. 2: Foto do FunCubeDongle Pro+

A Figura 3 é uma captura de tela do software Spark configurado para a transmissão DRM.

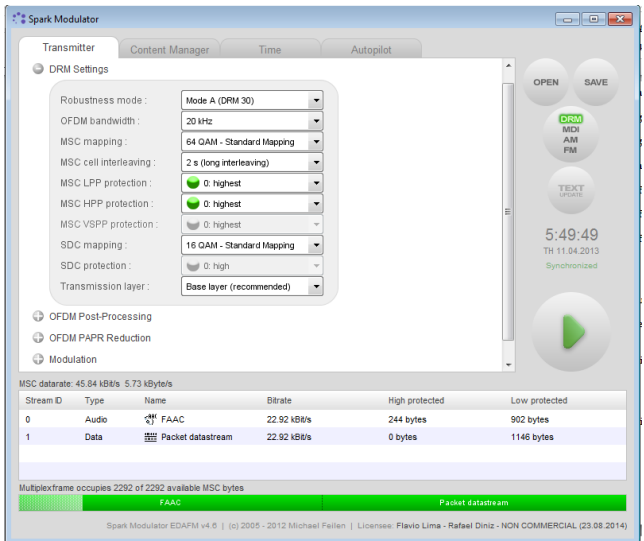

Fig. 3: Captura de tela do software Spark

E a Figura 4 mostra o receptor Dream recebendo a transmissão DRM com a aplicação Ginga.

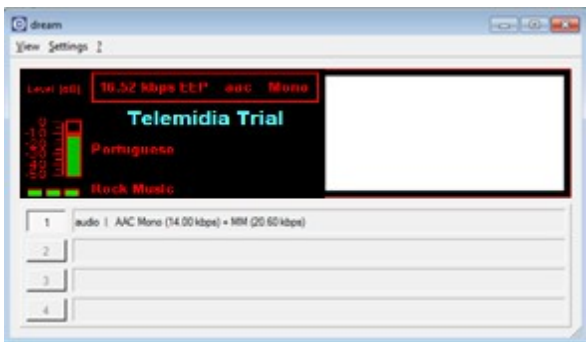

Fig.4: Tela do receptor Dream recebendo um sinal DRM que estava carregando um aplicativo Ginga e o áudio

O Dream com sucesso gravou o aplicativo Ginga retirado do carrossel no disco, e um script desenvolvido para o teste, ao perceber que o aplicativo foi recebido, executou-o com o exibidor Ginga (aplicativo e script disponível em [17]).

Na figura 5 é mostrada uma tela da aplicação, que consiste em um menu com opções de acesso à informações da Empresa Brasil de Comunicação e da programação da Rádio Nacional da Amazônia.

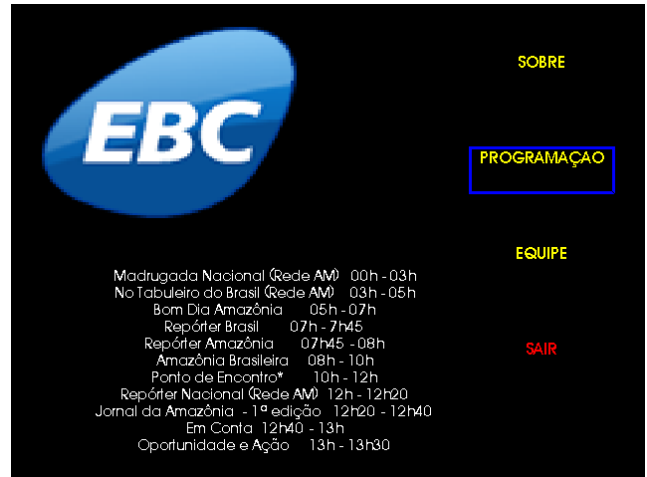

Fig. 5: Tela da aplicação sendo executada pelo exibidor Ginga automaticamente após ter sido recebido pelo receptor DRM Dream

Com relação ao tempo que o aplicativo demorou para ser transmitido, recebido e executado, segue uma breve discussão. Os arquivos que compõem o aplicativo, sem compactação, apresentaram os seguintes tamanhos, em bytes:

| demo1.ncl | 5734 |
|---|---|
| equipe.txt | 7 |
| logo.png | 6922 |
| prog.txt | 14 |
| sair.txt | 5 |
| sobre.txt | 6 |
| text1.txt | 537 |
| text2.txt | 794 |
| text3.txt | 122 |
| TOTAL | 14141 |

No entanto, como a maioria dos arquivos contém texto, com exceção do "logo.png", e texto normalmente consegue ser bastante compactado, foi utilizado a ferramenta "tar" para agrupar os arquivos e a ferramenta "xz", que implementa o algoritmo LZMA2, para compactar os arquivos. O tamanho final da aplicação, já compactada foi:

| app.tar.xz | 8724 |
|---|---|

Como no teste realizado foram alocados aproximadamente 5000 bits/s para transmitir a aplicação, temos que o valor esperado para a transmissão da aplicação completa foi:

8724 bytes * 8bits/byte = 69792 bits

69792 bits / 5000bits/s = 13,9584 s

O tempo estimado foi portanto aproximadamente 14s.

Empiricamente percebemos que após o receptor Dream iniciar a decodificação do sinal DRM, após aproximadamente 18s a aplicação já estava sendo executada. O tempo obtido foi um pouco mair do que o tempo esperado devido aos possíveis overheads de bits na multiplexação e de tempo para o exibidor Ginga carregar.

## 5. ANÁLISE DOS DADOS E PERSPECTIVAS PARA O RÁDIO DIGITAL BRASILEIRO

Os dados obtidos com o experimento nos mostram que é possível a transmissão e recepção de aplicações interativas mesmo em um canal de rádio digital operando a uma baixa taxa de bits, ou seja, o Sistema Brasileiro de Rádio Digital pode ter interatividade utilizando o middleware Ginga.

Vale ressaltar que o meio rádio está em funcionamento no Brasil a mais de 90 anos, desde 1922[18], tendo sido inaugurado oficialmente em ocasião do I Centenário da Independência do Brasil.

O Rádio é o único meio de comunicação que cobre 100% do território brasileiro através das transmissões em Ondas Médias, Ondas Tropicais, Ondas Curtas e na faixa do VHF. Desde as fronteiras amazônicas até grandes capitais, sinais de rádio sempre podem ser captados. A Rádio Nacional da Amazônia, por exemplo, cobre mais da metade do território nacional e alguns países das Américas que estão na direção do sistema irradiante da emissora, que fica em Brasília com antenas apontadas para o norte[19].

O Rádio Digital traz um grande um grande avanço com relação à qualidade do áudio analógico, principalmente no caso de emissoras que utilizam modulação em amplitude nas bandas de OM, OT e OC, além da possibilidade da multiprogramação, surround sound 5.1 e interatividade.

A interatividade no rádio digital pode ter significativa relevância tomando-se em consideração que regiões remotas do país poderão ter acesso à conteúdos de diferentes naturezas como os educativos, com informações de serviços públicos e de interesse social.

Pelo fato da radiodifusão ser um serviço essencial, no qual as emissoras com concessão devem obrigatoriamente estar no ar, um exemplo de uso crítico do Ginga seria o de informar rotas de saída, por exemplo, em casos de catástrofes naturais, auxiliando o sistema EWS (Emergency Warning System).

Um uso relevante do Ginga em emissoras comerciais, por exemplo, seria o de potencializar seus anúncios de forma visual, visto que o rádio tradicionalmente somente possui conteúdo de áudio somente.

É notável que muitos ouvintes de rádio hoje usam aparelhos de telefone celular para receber os sinais da estação, ou aparelhos automotivos. Nesses casos em que o receptor já possui uma boa tela embutida, aplicativos Ginga poderão ter praticamente todas as funcionalidades que já tem na TV Digital com a possibilidade de não cobrir nenhum conteúdo inerente do meio, visto que se o aplicativo tiver somente elementos visuais como imagens e textos, não existe interseção do conteúdo primário (áudio) com o conteúdo do aplicativo Ginga.

Além disso muitos receptores de rádio, como os presentes em celular e os automotivos, estão conectados à receptores GPS, o que permite ao aplicativo interativo escolher um conteúdo direcionado ao ouvinte de determinada localidade para, por exemplo, exibir uma informação de trânsito, propaganda de algum estabelecimento nas redondezas ou mesmo chavear o fluxo de áudio.

## 6. REFERÊNCIAS